# Production of $^{26}$Al, $^{44}$Ti and $^{60}$Fe in Supernovae- sensitivity to the helium burning rates

**Clarisse Tur**

*NSCL, Michigan State University and JINA*
*1 Cyclotron, East Lansing MI 48824-1321, USA*

**Alexander Heger**

*School of Physics and Astronomy, University of Minnesota, and JINA*
*Twin Cities, Minneapolis, MN 55455-0149, USA*
*E-mail:* `alex@physics.umn.edu`

**Sam M. Austin[1]**

*NSCL, Michigan State University and JINA*
*1 Cyclotron, East Lansing MI 48824-1321, USA*
*E-mail:* `austin@nscl.msu.edu`

We have studied the sensitivity of supernova production of the gamma emitting nuclei $^{26}$Al, $^{44}$Ti and $^{60}$Fe to variations of the rates, $R_{3\alpha}$ and $R_{\alpha 12}$, of the triple alpha and $^{12}$C($\alpha,\gamma$) reactions. Over a range of twice their experimental uncertainties we find variations in the production of $^{60}$Fe by more than a factor of five. Smaller variations, about a factor of two to three, were observed for $^{26}$Al and $^{44}$Ti. The yields of these isotopes change significantly when the abundances of Lodders (2003) are used instead of those of Anders and Grevesse (1989). These sensitivities will limit conclusions based on a comparison of observed gamma ray intensities and stellar models until the helium burning rates are better known. Prospects for improving the helium burning rates are discussed and a new version of the Boyes rate for $^{12}$C($\alpha,\gamma$) is presented.

---

[1] Speaker

# 1. Introduction

Astronomical observations of gamma rays from radioactive nuclei in the cosmos provide information about the sites and nature of stellar nucleosynthesis that is difficult to obtain in other ways. For example, the observed intensity of $^{26}$Al gamma rays can be used to estimate the galactic supernova rate, and the ratio of $^{26}$Al and $^{60}$Fe gamma-ray fluxes provides a test of supernova models. Such comparisons, however, depend on the robustness of supernova model predictions to changes in the nuclear reaction rates within their experimental uncertainties. The rates of the helium burning reactions $R_{3\alpha}$ and $R_{\alpha 12}$ are particularly important; their values affect the relative production of $^{12}$C and $^{16}$O which in turn affects all later stages of stellar evolution. These rates are not well known, with uncertainties of ±12% and ±25%. In this paper we describe a systematic study of the changes in supernova synthesis of $^{26}$Al, $^{44}$Ti and $^{60}$Fe that result from changes in the helium burning reaction rates and in the initial stellar abundances.

# 2. Procedures

We used the KEPLER code to simulate the evolution of 15, 20, and 25 $M_{sun}$ stars from hydrogen burning up to core collapse. A piston placed at the base of the oxygen shell and depositing a kinetic energy of 1.2 Bethe to the reaction products was then used to simulate the supernova explosion. For a more detailed description of the models see [1, 2], and for an overview of the reaction rates used see [2, 3]. Fallback and mixing were estimated following [4]. The central values of $R_{3\alpha}$ and $R_{\alpha 12}$ were taken from Caughlan and Fowler [5] and 1.2 times the rate of Buchmann [6], with uncertainties of ±12% and ±25%, resp. In two series of models, $R_{3\alpha}$ was varied over a range of ±2σ while holding $R_{\alpha 12}$ at its central value and $R_{\alpha 12}$ was varied in a range of ±2σ while holding $R_{3\alpha}$ at its central value. Each calculation was repeated for the Anders-Grevesse (AG89) [7] and the Lodders (Lo03) [8] abundances. Although not discussed in this paper, calculations were also performed in which $R_{3\alpha}$ and $R_{\alpha 12}$ were simultaneously varied, keeping their ratio constant. In total, over 200 evolution models were calculated.

# 3. Results

For a comparison to earlier calculations and references see [2]. In Fig. 1 we show a sample of the present results. In summary: (1) For $^{60}$Fe, variations of a factor of 4.4 (7.8) are seen for a ±σ (±2σ) range of rates. For $^{26}$Al the comparable numbers are 1.5 (2.8) and for $^{44}$Ti 1.2 (2.2). More detailed results are shown in Fig. 2, where the quadrature sums of the differences for $R_{3\alpha}$ and $R_{\alpha 12}$ are also shown. (2) The differences resulting from variations of $R_{3\alpha}$ and $R_{\alpha 12}$ are comparable in size. (3) The results, especially for $^{60}$Fe whose yield varies by a factor of two, are sensitive to the abundance differences of AG89 and Lo03. (4) The variations are somewhat smaller when averages over the three stars are considered, but remain large, especially for $^{60}$Fe, as shown in Fig. 1.

Uncertainties in other rates and in the model (e.g., convection) details further increase the yield uncertainties.

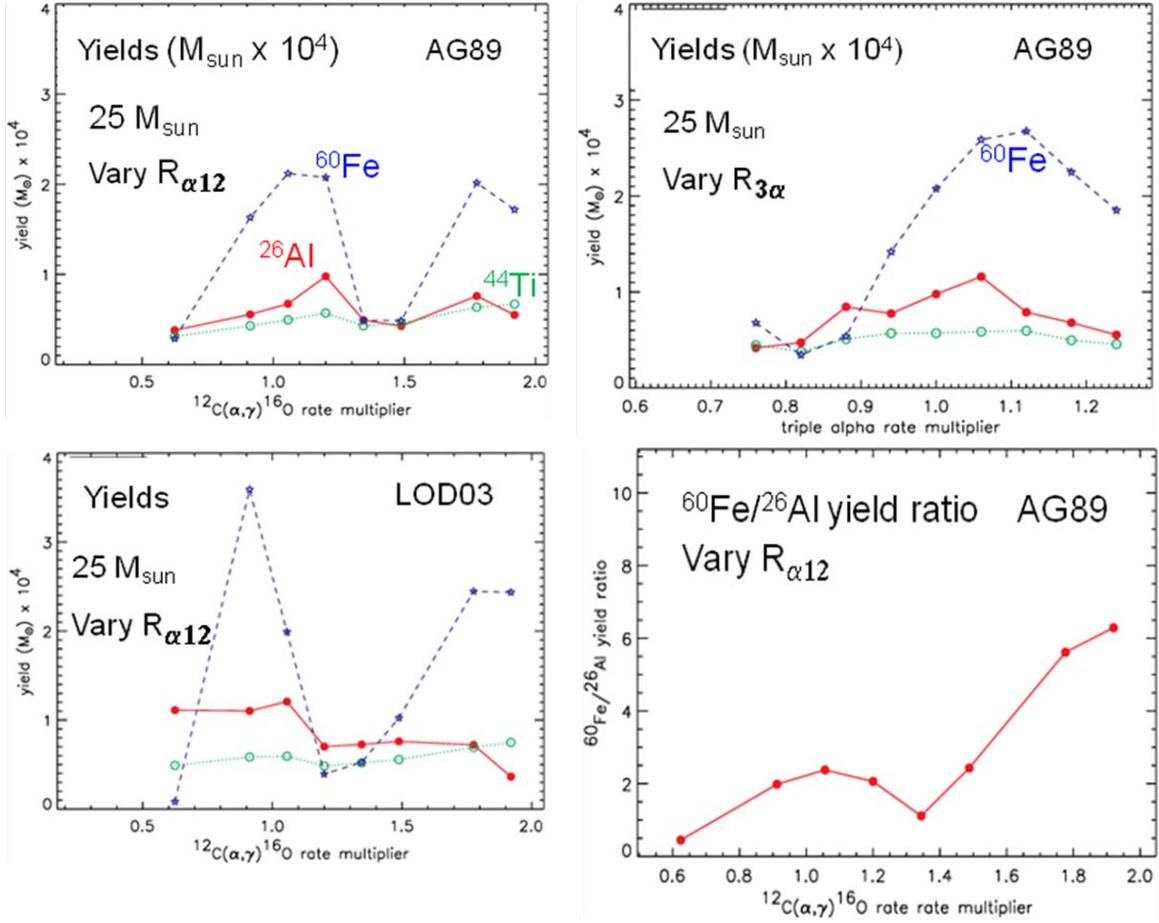

Fig. 1: (Upper left) Changes of the yields for a 25 $M_{sun}$ star when $R_{\alpha 12}$ is varied over a ±2σ range. (Upper right) Same for $R_{3\alpha}$. (Lower left) Changes when $R_{\alpha 12}$ is varied for the Lod03 abundances. (Lower right) Three star average of changes in the ratio of $^{60}$Fe and $^{26}$Al yields when $R_{\alpha 12}$ is varied.

| Rate | $^{26}$Al(1σ) | $^{44}$Ti(1σ) | $^{60}$Fe(1σ) | $^{26}$Al(2σ) | $^{44}$Ti(2σ) | $^{60}$Fe (2σ) |
|---|---|---|---|---|---|---|
| $R_{3\alpha}$ | 1.5 | 1.2 | 5.0 | 2.8 | 1.6 | 7.8 |
| $R_{\alpha 12}$ | 2.3 | 1.3 | 4.4 | 2.6 | 2.2 | 7.4 |
| Σ | 2.7 | 1.8 | 6.6 | 3.8 | 2.7 | 10.7 |

Fig. 2: Ratio of maximum and minimum yields for ranges of ±σ and ±2σ variations in $R_{3\alpha}$ and $R_{\alpha 12}$ for a 25 $M_{sun}$ star.

Fig. 2 shows, for example, that over a range of ±σ for $R_{3\alpha}$, variations of a factor of 5.0 are found for $^{60}$Fe. The quadrature sums of the results for $R_{3\alpha}$ and $R_{\alpha 12}$ are also given in the Σ row, as a rough estimate of the total uncertainty resulting from variations in $R_{3\alpha}$ and $R_{\alpha 12}$.

These results affect the use of gamma ray observations to constrain the galactic supernova rate and to test supernova models. By comparing supernova yield estimates and observed gamma intensities, Diehl et al. [9] argue that the $^{26}$Al gamma flux corresponds to 1.9 ± 1.1 supernova events/century. Given the $^{26}$Al uncertainties presented here, it seems that the error is larger. Wang et al. [10] observe a yield ratio $^{60}$Fe/$^{26}$Al = (60/26)(0.15 ± 0.06). Comparisons with supernova model calculations will be unconvincing because of the large yield uncertainties corresponding to the uncertainties in the helium burning rates.

**4. Improving the helium burning reaction rates.**

Given the nucleosynthesis uncertainties associated with the present uncertainties in $R_{3\alpha}$ and $R_{\alpha 12}$ it is important to examine the prospects for decreasing their uncertainties. The rate of the triple alpha process is proportional to the radiative width of the 7.6 MeV $0^+$ state in $^{12}$C-the Hoyle state. This width is determined experimentally from measurements of the three quantities indicated by arrows below.

$$R_{3\alpha} \propto \Gamma_{rad}^{Hoyle} = \frac{\Gamma_\gamma + \Gamma_\pi}{\Gamma} \frac{\Gamma}{\Gamma_\pi} \Gamma_\pi$$
$$\hspace{3cm} \Uparrow \hspace{1cm} \Uparrow \hspace{0.5cm} \Uparrow$$
$$\hspace{3cm} (A) \hspace{0.5cm} (B) \hspace{0.3cm} (C)$$

The radiative branch (A) is known (from 8 measurements) to ±2.7 %, the pair branch (from three measurements) to ±9% and the pair width (from 2 measurements) to ±6% giving an overall uncertainty of about ±12%. A recent measurement [11], however, has reduced the uncertainty in (C) to ±3.2%. And an experiment by a WMU+MSU+ANL group [12] aims to determine the small pair branch (7 x10$^{-6}$) to within ±4%. If this is achieved, the overall uncertainty will be halved to ±6%.

The situation for $R_{\alpha 12}$ is less clear. In spite of heroic efforts, both the strength and energy dependence of the reaction remain poorly known. As a result, many theoretical calculations use the unpublished effective interaction obtained by Boyes et al. [4]. Boyes proceded by fixing $R_{3\alpha}$ at its central value, assuming an energy dependence for $R_{\alpha 12}$ as we have done, and then adjusting the normalization of $R_{\alpha 12}$ so as to minimize the spread in production factor for a set of light and medium weight nuclei. The result was a value 1.2 ± 0.1 times that quoted by Buchman [6].

We have attempted to validate and improve that result with the same procedure, but using a larger sample of stars and taking into account processing in the supernova explosion, ignored by Boyes. The results shown in Fig. 3 give a value of 1.3, in surprisingly good agreement with Boyes, surprising because the explosive processing ignored by Boyes changes many abundances by a factor of two. The uncertainty of ±0.1, however, taken from Boyes and shown in Fig. 3, seems too small, espcially given the shift of +0.1 and the large rms at the central value. A more conservative estimate is 1.3 ± 0.25 times Buchmann [6].

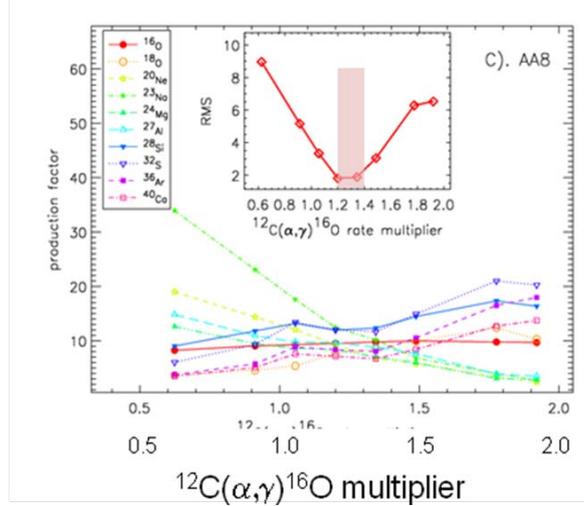

Fig. 3: Average Production Factors for the light and medium weight nuclei shown and a set of eight stars with masses from 13 to 27 $M_{sun}$. The inset shows the rms variation about the average. The uncertainty band is taken from Boyes.

A more important criticism is that this is an effective interaction and should probably only be used in the situation for which it was derived: KEPLER and its modelling assumptions, central $R_{3\alpha}$, and AG89 abundances. It may not be valid for other mass ranges or metallicities. For example when the procedure is applied to the Lod03 abundances there is not a well defined minimum [1].

**5.Summary**

The rates for the triple alpha and $^{12}C(\alpha, \gamma)$ reactions are not sufficiently well know to permit reliable estimates of the production of $^{26}$Al, $^{44}$Ti and $^{60}$Fe in supernovae. Prospects for improved triple alpha rates are good, but the situaion is less clear for $^{12}C(\alpha, \gamma)$.

**References**

[1]  C. Tur, A. Heger, and S. M. Austin, *Astrophys. J.* **671**, 821 (2007); **702**, 1068 (2009).
[2]  C. Tur, A. Heger, and S. M. Austin, *Astrophys. J.* **718**, 357 (2010).
[3]  T. Rauscher, A. Heger, R. D. Hoffman, and S. E. Woosley, *Astrophys. J.* **576**, 323 (2002)
[4]  S. E. Woosley and A. Heger, *Phys. Rep.* **442** 269 (2007)
[5]  G. R. Caughlan and W. A. Fowler, *At. Data Nucl. Data Tables* **40**, 283 (1988).
[6]  L. R. Buchmann, *Astrophys. J.* **468**, L127 (1996); errata: **479**, L153 (1997).
[7]  E. Anders and N. Grevesse, *Geochim. Cosmochim. Acta* **53**, 197 (1989)
[8]  K. Lodders, *Astrophys. J.* 591, 1220 (2003)
[9]  R. Diehl, H. Halloin, and K. Kretschmer, *Nature* **439**, 45 (2006)
[10] W. Wang, *et al, Astron. Astrophys.* **469**, 1005 (2007)
[11] M. Chernykh, *et al.* .*Phys. Rev. Lett.* **105**, 022501 (2010)
[12] A. Wuosmaa, *et al.*, Private Communication